\documentclass{mn2e}
\usepackage{epsfig}

\def\msun{{\rm M_{\odot}}}

\title[Black Hole Blackbodies]
{Black Hole Blackbodies}
\author[A.R.~King, E.M.~Puchnarewicz]{
A.R.~King$^1$, E.M.~Puchnarewicz$^2$\\
1. Theoretical Astrophysics Group, University of Leicester,
Leicester, LE1~7RH, UK\\
2. Mullard Space Science Laboratory University College, London Holmbury
St. Mary, Dorking, Surrey RH5 6NT }

\begin{document}

\maketitle

\begin{abstract}
Many black--hole sources emit a substantial fraction of their
luminosities in blackbody--like spectral components. It is usual to
assume that these are produced in regions at least comparable in size
to the hole's Schwarzschild radius, so that a measure of the emitting
area provides an estimate of the black hole mass $M$. However there is
then no guarantee that the source luminosity (if isotropic) obeys the
Eddington limit corresponding to $M$. We show that the apparent
blackbody luminosity $L_{\rm sph}$ and temperature $T$ must obey the
inequality $L_{\rm sph} < 2.3\times 10^{44}(T/100\ {\rm eV})^{-4}$
~erg~s$^{-1}$ for this to hold. Sources violating this limit
include ultrasoft AGN and some of the ultraluminous X--ray sources
(ULXs) observed in nearby galaxies. We discuss the possible
consequences of this result, which imply either super--Eddington or
anisotropic emission in both cases. We suggest that the ultrasoft AGN
are the AGN analogues of the ULXs.
\end{abstract}


\section{Introduction}

Many luminous accreting sources have prominent spectral components
which appear approximately blackbody. If the accretor is a black hole
it is natural to assume that the dimensions of the emitting surface
are comparable to the Schwarzschild radius, as this is the region
where most of the gravitational potential energy is released. However
the result of this procedure is not necessarily compatible with the
Eddington limit on the source luminosity, derived from the requirement
that the radiation pressure should not drive the accreting matter
away. Examples of this for quasars have appeared in the literature
(Puchnarewicz, 1994; Molthagen, Bade and Wendker 1998; Puchnarewicz
1998). Here we investigate the question systematically, and show that
various classes of accreting black holes do not straightforwardly
correspond to sub--Eddington blackbodies. In general the resolution of
this difficulty appears to require super--Eddington or anisotropic
emission.

\section{Blackbodies}

The luminosity and effective temperature of an optically thick emission
region of typical size $R$ near a black hole are related by 
\begin{equation}
L = 4\pi p\sigma R^2T^4 \label{bb}
\end{equation}
where the factor $p$ allows for deviations from isotropy and spherical
symmetry in both source geometry and emission pattern. We write $R =
rR_s$, where $R_s = 2GM/c^2$ is the Schwarzschild radius, so that
\begin{equation}
L = {16\pi\sigma G^2\over c^4}pr^2M^2T^4. \label{bbm}
\end{equation}
We can also express $L$ in terms of the Eddington value $L_{\rm Edd}$, as
\begin{equation}
L = l{4\pi GMm_pc\over \sigma_T}, \label{edd}
\end{equation}
with $l = L/L_{\rm Edd}$.  If the emission is anisotropic we must be
careful to distinguish between the true luminosity $L$ and the
`spherical' value $L_{\rm sph}$ assigned by a distant observer. As the
translation from observed flux to luminosity generally assumes
isotropic emission these are related by
\begin{equation}
L = bL_{\rm sph}. \label{beam}
\end{equation}
Here $b \la 1$ if (as expected in a flux--limited sample) the
observer lies within the `beam' of the source (note that `beaming'
here merely means `anisotropic', and does {\it not} necessarily imply
relativistic beaming). Eliminating $M$ between (\ref{bbm},
\ref{edd}) and using the definition of $L_{\rm sph}$ we get
\begin{equation}
L_{\rm sph} = {L\over b} = {2.3\times 10^{44}\over T_{100}^4}{l^2\over
pbr^2}\ {\rm erg\ s^{-1}}, \label{lim1}
\end{equation}
and thus
\begin{equation}
M = {1.8\times 10^6\over T_{100}^4}{l\over pr^2}\msun, \label{m1}
\end{equation}
where $T_{100}$ is $T$ in units of 100eV.  Hence any object which
violates the limit
\begin{equation}
L_{\rm sph} < L_0 = {2.3\times 10^{44}\over T_{100}^4} \label{lim2}
\end{equation}
must have $l^2/pbr^2 > 1$, i.e. must either be super--Eddington ($l >
1$), or emit from a region much smaller than the Schwarzschild radius
($r<1$), or emit anisotropically ($pb < 1$). In the latter case, the
luminosity $L$ will be {\it less} (by a factor up to $b$) than
estimated at infinity by assuming isotropic emission. Given a choice
of $l, p, r$ we get a mass estimate from (\ref{m1}). A simple
interpretation of the limit (\ref{lim2}) is that it ensures that the
lower mass bound given by respecting the Eddington luminosity (cf
eq. \ref{edd}) is compatible with the upper bound (cf eq. \ref{m1})
implied by the blackbody temperature, assuming isotropic emission from 
a region of size comparable with the Schwarzschild radius ($pb = r = 1$).

We can generalise this result to the case where the blackbody emission
is not the only luminosity component detected, i.e., where (\ref{bb})
is replaced by
\begin{equation}
L = fL_1 + 4\pi p\sigma R^2T^4, \label{dbb}
\end{equation}
where $f$ is again a `beaming' factor reflecting the difference
between the real luminosity $fL_1$ and the value $\leq L_1$ calculated
by a distant observer assuming isotropic emission. For example, $fL_1$
might be the power--law continuum frequently observed. Then using $R=
rR_s$ and (\ref{edd}) as before we get
\begin{equation}
fL_1 = -{pr^2\over l^2L_0}L^2 + L \label{dlim}
\end{equation}
Since $L_1 >0$ we have
\begin{equation}
L < L_0{l^2\over pr^2} \label{dlim1}
\end{equation}
and maximizing the rhs of (\ref{dlim}) as $L$ varies implies 
\begin{equation}
L_{\rm 1, sph} = b^{-1}L_1 < L_0{l^2\over 4fpbr^2}. \label{L_1lim}
\end{equation}
Hence if (\ref{lim2}) is violated, where $L_{\rm sph}$ is now
interpreted as the total luminosity, not that of the blackbody
component alone, we can draw the same conclusions as before, i.e. that
the whole source is either super--Eddington, or the blackbody
component comes from a region smaller than the Schwarzschild radius or
is emitted anisotropically. The main importance of this result is the
implication that our conclusions above for the blackbody component
alone (and the corresponding mass) are not affected by inaccuracies in
subtracting a diluting non--blackbody component. The new feature is
that these same conclusions must hold (with the possible alternative
of beaming of the diluting continuum, i.e. $f < 1$) if the
non--blackbody emission appears to violate the limit
\begin{equation}
L_{\rm 1, sph} < 0.25L_0 = {6\times 10^{43}\over T_{100}^4}\ {\rm erg\
s^{-1}}.
\end{equation}
In principle this gives a formally tighter mass limit
\begin{equation}
M < {4.4 \times 10^5\over T_{100}^4}{l\over pr^2} \msun.
\end{equation}
The main restriction on using these limits more widely is of course the
difficulty of making accurate temperature determinations if the
dilution is severe.

\section{`Super--Eddington' sources}

The condition (\ref{lim2}) divides sub--Eddington sources from those
where more complex phenomena occur.  Sources respecting this limit can
be sub--Eddington; sources above it are either super--Eddington, or
emit from a surface smaller than the black hole horizon, or emit
anisotropically. Fig.~1 shows the limit plotted for a range of
luminosities and temperatures covering both stellar--mass and
supermassive systems.

\subsection{Stellar--mass black holes}

For stellar--mass systems it is convenient to rewrite (\ref{lim1},
\ref{m1})) as
\begin{equation}
L_{\rm sph} = 1.4\times 10^{39}\biggl({T\over 2\ {\rm
keV}}\biggr)^{-4}{l^2\over pbr^2}\ {\rm erg\ s}^{-1} \label{lgal}
\end{equation}
and
\begin{equation}
M = 10\biggl({T\over 2\ {\rm keV}}\biggr)^{-4}{l\over pr^2}\msun
\label{mgal}.
\end{equation}
These forms agree with the well--known facts that X--ray bursters
($L_{\rm sph} \la 2\times 10^{38}$ erg s$^{-1}$, $T \la 2$~keV)
radiate isotropically ($pb \simeq 1$) at close to the Eddington limit
($l \simeq 1$) and are neutron stars ($r \simeq 3$) with mass $M \la
1.4\msun$. Outbursting soft X--ray transients reach luminosities
$L_{\rm sph} \ga 10^{39}$~erg~s$^{-1}$, but are generally ultrasoft,
and so do not threaten the limit (\ref{lim2}). They are therefore
consistent with the Eddington limit for black--hole masses $\sim 5 -
10 \msun$. Some of the ULX sources (see below) which appear to violate
the limit may however be unrecognised transients.

\subsection{Ultraluminous X--ray sources (ULXs)}

The ULXs are by definition a group of non--nuclear sources in nearby
galaxies where the solar--mass Eddington limit is apparently violated,
i.e. $L_{\rm sph} \la 10^{38}$\ erg\ s$^{_1}$ (e.g. Makishima et al.,
2000). We see from Fig.~1 that the ULXs straddle the line $L_{\rm sph}
= L_0$, potentially allowing an interpretation as sub--Eddington
sources in some cases. However for $T \la 2$~keV the required
black--hole mass exceeds $10\msun$ (cf eq. \ref{m1}). This led
initially to claims that these objects had masses intermediate between
stellar values and supermassive ones. However it is hard to reconcile
this idea with the clear link with recent star formation shown by {\it
Chandra} observations of the Antennae (Fabbiano, Zezas \& Murray,
2001) which reveal $\ga 10$ ULXs.  A more likely possibility is that
most (if not all) ULXs involve beamed, and possibly transient,
emission from a shortlived luminous phase of massive X--ray binary
evolution (King et al., 2001). In any case, several of these sources
violate the limit (\ref{lim1}) as their temperatures are too high for
the assigned `intermediate' masses (cf eq. \ref{mgal}). Recognising
this, Makishima et al (2000) suggested that their accretion discs
extended very close to the horizon of a maximally rotating Kerr black
hole ($r \sim 0.5$). Given the errors this may make all ULXs compatible
with the $L_{\rm sph} < L_0$ constraint, although only marginally.

\subsection{Ultrasoft quasars}

The soft X--ray excess in AGN (i.e. an excess of $\sim 0.1-0.5$~keV
flux above an extrapolation of the $\sim 2-10$~keV continuum) is
believed to be the high energy tail of an accretion disc spectrum
(Arnaud et al. 1985; Pounds et al., 1987; Turner \& Pounds 1989;
Masnou et al. 1994). A spectral turnover in soft X--rays that would
fix an effective temperature for an accretion disc is rarely observed
however, implying relatively cool discs. An upper limit on the disc
temperature can be derived from the slope of the soft X--ray excess,
and a lower limit from constraints provided by the UV continuum.
Using combined IPC and MPC data from {\sl Einstein}, Urry et
al. (1989) fitted blackbodies of less than 10~eV to the soft X-ray
components of their sample of AGN. By combining their PG quasar soft
X--ray sample with the UV results of Zheng et al. (1997), Laor et
al. (1997) found that the overall UV--soft X--ray continuum of quasars
peaked at around 1000~\AA\, again corresponding to $T \sim 10$~eV. The
typical luminosities $\sim~10^{44}$ to $10^{45}$~erg~s$^{-1}$ show
that the PG sample quasars are easily interpreted as sub--Eddington
($l < 1$), isotropic ($pb =1$) emitters with masses $\la$ a few
$\times 10^9\msun$ (cf (\ref{lim1}, \ref{m1}); see also Fig. 1).

\begin{figure}
  \begin{center}
    \epsfig{file=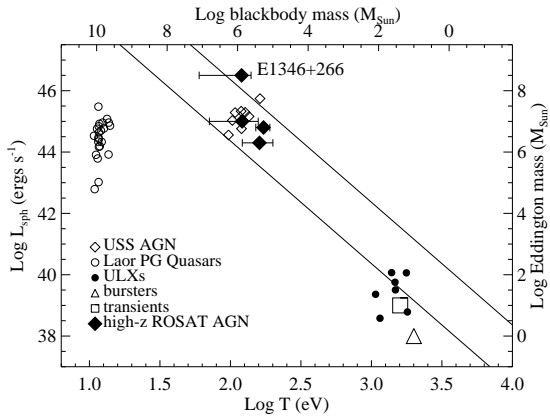, width=8cm}
  \end{center}
\caption{ Luminosity and blackbody temperature for bright X--ray
blackbody sources. 
Filled diamonds are confirmed ultrasoft AGN. E1346+266 is labelled
separately and errors on the blackbody temperature are 99~percent. The
other three are {\sl ROSAT}-selected ultrasoft AGN taken from
Puchnarewicz (1998): errors on the blackbody temperatures are
90~percent. Open diamonds are the remaining high-z USS AGN.  
Also plotted individually are the Laor et al. sample of
PG quasars (crosses) and ULXs (stars). X--ray transients (square) and
bursters (triangle) are shown schematically only, in the interests of
clarity. The solid line is the limit $L_{\rm sph}=L_0$. Sources below
this line are compatible with the Eddington, isotropy and size
constraints $l < 1, pb = r =1$, but must have masses $M$ respectively
above and below the values given on the rh vertical and upper
horizontal scales. The thin line represents a luminosity 100
times this limit.
}
\end{figure}

AGN with hotter discs must have lower luminosities to remain
sub--Eddington. RE~J1034+396 is a rare AGN that does have a
measureable high-energy turnover in soft X-rays, with $T \simeq
120$~eV. Its luminosity $L_{\rm sph} \sim 10^{43}$ erg s$^{-1}$ is
consistent with sub-Eddington accretion, and a mass $M \sim
7\times10^6 \msun$ (Puchnarewicz et al. 2001). This is consistent with
an earlier claim by Pounds, Done \& Osborne (1995) that RE~J1034+396
is a supermassive analogue of stellar--mass black holes in their high
state. 

However, there are several examples of AGN whose accretion disc
temperature is as hot that in RE~J1034+396, but with
much higher luminosities. The ultrasoft X-ray component in
E1346+266 (z=0.92) was confirmed by the {\sl ROSAT} PSPC to have a
rest-frame temperature of 120$^{+20}_{-60}$~eV (errors are a
conservative 99~percent;
Puchnarewicz, Mason \& Cordova 1994). With an intrinsic luminosity of
$\sim5\times10^{46}$~erg~s$^{-1}$ (assuming unbeamed, spherically
symmetric emission), this corresponds to more than 100 times the
Eddington limit. The three AGN from Puchnarewicz
(1998) have rest-frame blackbody temperatures of $\sim$150~eV and
luminosities between 2$\times10^{44}$~erg~s$^{-1}$ and
10$^{45}$~erg~s$^{-1}$; the order of 10 
times ``super-Eddington'' (see Fig. 1). A
further eight hot (kT$_{\rm BB}\sim$100~eV), high-luminosity 
($L_{\rm sph}
\sim3\times10^{44}-10^{46}$ erg s$^{-1}$) candidates have been 
identified as part
of the {\sl Einstein} UltraSoft Survey (USS; Puchnarewicz et
al. 1992; C\'ordova et al. 1992; Thompson \& C\'ordova, 1994), 
although further data would be required to confirm the soft spectra
indicated by {\sl Einstein}. 

There has been evidence to suggest that the velocities of the broad
line regions (BLRs) in AGN may provide an alternative way of measuring
the black hole mass (see eg. Laor 1998). However, four of the six
high-luminosity quasars discussed in Puchnarewicz (1998) have slow
BLRs, which would seem to refute this relationship. A simple way of
addressing this is to invoke anisotropy in the continuum emission,
which would reduce the black hole mass and the broad line region
velocity. However, this involves several naive assumptions about
the relation between the black hole mass, accretion disc spectrum and the
conditions, geometry and velocity of the BLR. We caution that while
supporting our hypothesis of anisotropy in ultrasoft quasars, this
solution may not be as straightforward as it appears; we postpone
further discussion to future work. The more fundamental issue of
apparently super-Eddington emission from quasars remains, irrespective
of the conditions in the BLR.

\section{Discussion}

Fig.~1 shows that the ultrasoft AGN have luminosities $L_{\rm sph}$ a
factor $\sim 10 - 30$ above the value of $L_0$ appropriate to their
measured temperatures, while several ULX sources show a similar if
smaller effect. Before going further we should check if these results
could be spurious. Probably the most serious possible cause of error
arises from the claimed blackbody temperatures. These might be
systematically too high either because (a) the values of $T$ are
wrongly fitted, or (b) the spectra are not blackbody at all, but for
example the result of Comptonization, or the effects of electron
scattering in regions with low absorption opacity. Errors of the
right order (factors $\ga 3 - 10$) appear unlikely; and if the effect
is a physical one we have to explain why it occurs only in a subset of
AGN.  Nevertheless this area merits further study.

If we accept the values of $T$ appearing in Fig.~1, we should consider
the possibility that the emission may come from a region of total area
smaller than the Schwarzschild radius $R_s$, i.e. that $r < 1$. Since
most of the accretion energy is released in a region of order $R_s$,
this requires that it should be removed non--radiatively, and only
converted to radiation in a much smaller region. The only way of
ensuring this appears to involve magnetic fields. Merloni \& Fabian
(2001) have indeed suggested that magnetic reconnection in regions
comparable with the local disc thickness $H$ may be the primary
dissipation mechanism in accretion discs. However, they also require
that these regions should be triggered at heights at least an order of
magnitude larger than their size. Since $H \sim 0.1R$, heights $\ga R
\ga R_s$ are needed. In this picture, blackbody emission would result
from reprocessing the primary emission from these `lamppost' regions
on the disc surface. The reprocessing regions cannot be smaller than
the lamppost heights $\ga R_s$, implying $r > 1$ in this
model. Similar considerations probably apply in other magnetic energy
release pictures.

Thus assuming that the values of $T$ in Fig.~1 are not grossly in error,
and that the radiating regions of the sources are unlikely to be
significantly smaller than their Schwarzschild radii, we are left with
the alternatives that the sources with $L_{\rm sph} > L_0$ are
genuinely super--Eddington, or radiate anisotropically. We note that
either of these possibilities would probably remove the need for
intermediate masses $M \ga 100\msun$ in even those ULXs which do not
violate the limit $L_{\rm sph} > L_0$. Begelman (2002) has proposed a
mechanism allowing thin accretion discs to radiate at up to ten times
the Eddington limit, while King et al. (2001) have suggested that most
ULXs are anisotropic emitters.

The common feature here is that presumably both types of source are
supplied with mass at rates close to or above the Eddington value
$\dot M_{\rm Edd} \simeq L_{\rm Edd}/0.1c^2$. To zeroth order we would
expect the resulting accretion geometry to be similar in the two cases
despite the large difference in black hole mass. We therefore suggest
that the ultrasoft AGN are the supermassive analogues of the ULXs.

\section{Acknowledgments} 

Theoretical astrophysics research at Leicester is supported by a PPARC
rolling grant. We thank the referee, Ari Laor, for constructive comments.

\end{document}